\documentclass[pra,twocolumn,showpacs,preprintnumbers,amsmath,amssymb,superscriptaddress]{revtex4-1}
\usepackage{graphicx}% Include figure files
\usepackage{dcolumn}% Align table columns on decimal point
\usepackage{bm}% bold math
\usepackage[bookmarks, colorlinks, linkcolor=blue, citecolor=blue, urlcolor=blue]{hyperref}

\begin{document}
\title{Robust two-dimensional subrecoil Raman cooling\\ by adiabatic
transfer in a tripod atomic system}
\author{Vladimir S. Ivanov}\email{ivvl82@gmail.com}
\affiliation{Turku Centre for Quantum Physics, Department of Physics
and Astronomy, University of Turku, 20014 Turku, Finland}
\affiliation{Saint Petersburg State University of Information
Technologies, Mechanics and Optics, 197101 St.~Petersburg, Russia}
\author{Yuri V. Rozhdestvensky}
\email{rozd-yu@mail.ru}
\affiliation{Saint Petersburg State University of Information
Technologies, Mechanics and Optics, 197101 St.~Petersburg, Russia}
\author{Kalle-Antti Suominen}\email{Kalle-Antti.Suominen@utu.fi}
\affiliation{Turku Centre for Quantum Physics, Department of Physics
and Astronomy, University of Turku, 20014 Turku, Finland}
\date{\today}

\begin{abstract}
  We demonstrate two-dimensional robust Raman cooling in a four-level
  tripod system, in which velocity-selective population transfer is
  achieved by a \mbox{STIRAP} pulse. In contrast to basic 2D Raman
  cooling with square envelope pulses [Phys.~Rev.~A~\textbf{83},
  023407 (2011)], the technique presented here allows for a wide
  variation in the pulse duration and amplitude once the adiabaticity
  criterion is satisfied. An efficient population transfer together
  with attaining of a narrow spread of the resonant-velocity group
  leads to the narrowing of the velocity-distribution spread down to
  $0.1v_\mathrm{rec}$, corresponding to an effective temperature equal
  to $0.01 T_\mathrm{rec}$. This robust method opens new possibilities
  for cooling of neutral atoms.
\end{abstract}

\pacs{37.10.De}

\maketitle

\section{Introduction}

Control of the atomic degrees of freedom at low temperatures is the
starting point for many promising and popular research fields which
aim either at understanding and simulating the quantum nature of
particle collisions~\cite{Weiner1999}, photoassociation of atoms into
molecules~\cite{Calsamiglia2001,Jones2006}, many-body
effects~\cite{Bloch2008} and phase transitions~\cite{Dziarmaga2010},
or at applications such as quantum
computers~\cite{Nielsen2000,Stenholm2005} and atomic
clocks~\cite{Audoin2001,Ye2008,Derevianko2011}. Most effects are best
observed at ultralow temperatures, which can be nowadays achieved for
neutral atoms by a combination of laser cooling and subsequent
evaporative cooling~\cite{Cohen2011}. The latter process requires also
trapping of atoms, usually with a very tight confinement, which then
precludes efficient use of the method e.g.~for collimation of slow
atomic beams~\cite{Johnson1998,Theuer1999,Partlow2004} or in only
partially trapped systems such as one-dimensional optical
lattices~\cite{Lewenstein2012}. The latter situation is utilized e.g.
in optical atomic clocks~\cite{Ye2008,Derevianko2011}. Although tight
confinement and the subsequent discrete motional state structure for
atoms offers many methods for further cooling in a manner similar to
the cooling of trapped
ions~\cite{Morigi1997,Schmidt-Kaler2001,Leibfried2003} and open
possibilities for other interesting studies as quantum
computing~\cite{Cirac1995,Haffner2008} and
entanglement~\cite{Roghani2008,Roghani2011,Roghani2012}, alternative
approaches are needed in order to apply cooling at a more general
setting such as free space. This is the motivation for developing
further purely light-based methods for reaching similar temperatures
as with evaporative cooling, as discussed also in our previous work on
the topic~\cite{Ivanov2011,Ivanov2012}.

In the past, powerful cooling techniques have been designed to achieve
subrecoil temperatures of free atoms. The ``dark state''
cooling~\cite{Aspect1988} is very efficient but also limited to rather
collisionless situations (low densities). Raman cooling, on the other
hand, is not so density-limited, and deep subrecoil cooling in 1D has
been demonstrated~\cite{Kasevich1992} and extended to 2D and 3D
cooling~\cite{Davidson1994,Boyer2004} as well. The lowest temperature
in 2D, achieved for Cs atoms at NIST, Gaithersburg, is
0.15$T_\mathrm{rec}$~\cite{Boyer2004}, where $T_\mathrm{rec}$ is the
atomic recoil temperature. The suppression of further cooling is
associated with the required cumbersome setup of four Raman beam pairs
as well as limitations of the assumed $\Lambda$-type atomic state
system. Our recent suggestion of cooling in a tripod atomic level
system not only reduces the number of Raman beams by a factor of two,
but also allows one, in principle, to reach temperatures as low as
$0.01T_\mathrm{rec}$. However, more cooling cycles are required in 2D
Raman cooling as compared with 1D, which imposes strict demands on the
velocity precision of the Raman transfer~\cite{Ivanov2011}.

To overcome such a limitation, one can employ the robust transfer
process provided by \mbox{STIRAP}, as recently suggested by us for 1D
Raman cooling~\cite{Ivanov2012}. However, the process of transferring
atoms collected in the atomic ``dark state'' is not a trivial
extension of the 1D situation, and thereby the 2D case is of special
consideration. So far, \mbox{STIRAP} in a tripod system by resonant
laser beams has been experimentally explored only for atomic
beams~\cite{Theuer1999}, although far-off resonant lasers in general
are used for Raman cooling. This paper demonstrates theoretically 2D
Raman cooling by \mbox{STIRAP} going down to $0.01T_\mathrm{rec}$,
which nevertheless allows a wide variation in both the pulse envelope
and duration if only the adiabaticity criterion is satisfied. The
pulse duration needed for transfer exceeds the pulse durations for
normal Raman processes, so the advantage of robustness is attained
only if the cooling time is not a critical factor. This slowness
related to adiabaticity would restrict the application of the method
in atomic beam collimation to very slow beams. Another limitation
arises from the specific need for a tripod structure, which is not
present e.g. at the main transitions for the alkaline-earth
atoms~\cite{Machholm2001,Machholm2002} that are currently the
strongest candidate for optical atomic
clocks~\cite{Ye2008,Derevianko2011}.

The organization of this paper is as follows. The necessary atomic
tripod energy state diagram and the corresponding 2D laser beam
configuration are presented and discussed in Sec.~\ref{sec:tripod}. In
Sec.~\ref{sec:no_decay} we show that, as expected, large detuning from
the excited atomic state suppresses spontaneous decay, and the
resonant-velocity group of \mbox{STIRAP} under this condition is
discussed in Sec.~\ref{sec:res_group}. An efficient transfer of the
atoms from the original ground state, even in the case of such a large
detuning, leads to efficient 2D cooling, for which the parameters are
given in Sec.~\ref{sec:params}. The cooling itself is investigated in
Sec.~\ref{sec:cooling}, and our research is concluded by the summary
and discussion given in Sec.~\ref{sec:conclusion}.

\section{Tripod system and laser configuration}\label{sec:tripod}

\begin{figure}[htb]
(a)\\ \includegraphics{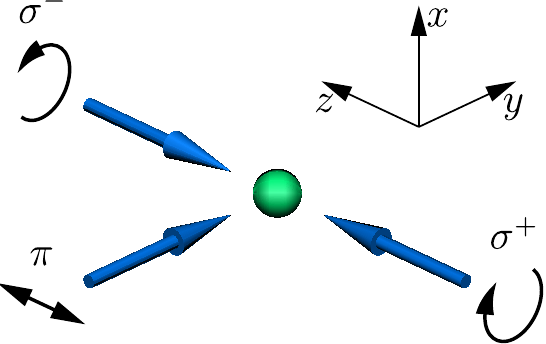}
\\
(b)\\ \includegraphics{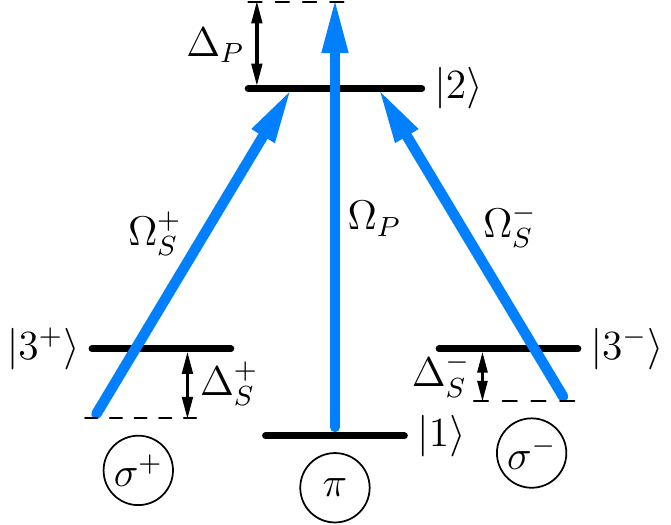}
\\
(c)\\ \includegraphics{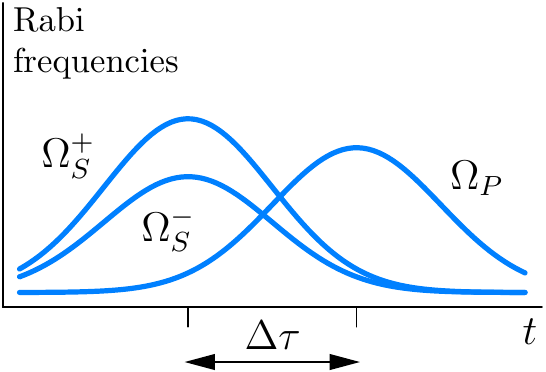}
\caption{\label{fig:tripod} (Color online) (a) The 2D laser
  configuration consists of three running waves, two
  circularly-polarized waves arranged along the $Oz$ axis and a
  $\pi$-polarized wave propagating in the $Oy$ direction. (b) The
  energy-level diagram of the tripod system coupled by laser beams.
  (c) Both Stokes pulses $\Omega_S^+(t)$ and $\Omega_S^-(t)$ forms a
  counterintuitive sequence with the pump pulse $\Omega_P(t)$ in order
  to carry out atomic population from state $|1\rangle$ to
  $|3^+\rangle$ and $|3^-\rangle$, respectively.}
\end{figure}

Consider a tripod system under conditions following closely metastable
Ne in Ref.~\cite{Theuer1999}. Pump laser couples state ($2p^53s$)
$^3P_0$ to an intermediate state ($2p^53p$) $^3P_1$ ($M=0$) which in
turn is coupled to magnetic substates $M=\pm 1$ of ($2p^53s$) $^3P_2$
by two Stokes lasers. The pump laser is a $\pi$-polarized running wave
propagating along axis $Oy$, and the Stokes lasers are
contra-propagating $\sigma$-polarized running waves arranged along the
$Oz$ axis (see Fig.~\ref{fig:tripod}(a)). Note that the metastable Ne
system is used only as an example. The classical electric field of all
three laser beams is written as
\begin{align}
  \mathbf E(\mathbf{r}, t) &= \frac{1}{2} \mathbf E_P e^{i\mathbf{k}_P
    \mathbf{r} - i\omega_P t} + \frac{1}{2} \mathbf E_S^+
  e^{i\mathbf{k}_S \mathbf{r} - i\omega_S^+ t}\nonumber
  \\
  &+ \frac{1}{2} \mathbf E_S^- e^{-i\mathbf{k}_S \mathbf{r} -
    i\omega_S^- t} + \text{c.c.}
\end{align}
The first term corresponds to pump laser with frequency $\omega_P$ and
wave vector $\mathbf{k}_P = k_P \mathbf{e}_y$; the two other terms
relatively correspond to $\sigma^+$- and $\sigma^-$-polarized Stokes
lasers with frequencies $\omega_S^+$, $\omega_S^-$ and wave vectors
$\mathbf{k}_S$, $-\mathbf{k}_S$, where $\mathbf{k}_S = k_S
\mathbf{e}_z$.

Figure~\ref{fig:tripod}(b) illustrates the atomic states coupled by
the laser configuration, with labelling
\begin{align}
\begin{aligned}
  &|1\rangle = |(2p^53s) \; ^3P_0, M=0\rangle,
  \\
  &|2\rangle = |(2p^53p) \; ^3P_1, M=0\rangle,
  \\
  &|3^+\rangle = |(2p^53s) \; ^3P_2, M=-1\rangle,
  \\
  &|3^-\rangle = |(2p^53s) \; ^3P_2, M=1\rangle.
\end{aligned}
\end{align}
Taking into account shifts in the centre-of-mass momentum, we consider
an atom of momentum $\mathbf{p}$ originally prepared in state $|1,
\mathbf{p}\rangle$. Then laser-atom coupling strengths are given by
\begin{align}\label{eq:coupling}
\begin{aligned}
  &\hat V |1, \mathbf{p}\rangle = \frac{\hbar}{2} \Omega_P(t)
  e^{-i\omega_P t} |2, \mathbf{p} + \hbar\mathbf{k}_P\rangle,
  \\
  &\hat V |3^\pm, \mathbf{p} + \hbar\mathbf{k}_P \mp
  \hbar\mathbf{k}_S\rangle = \frac{\hbar}{2} \Omega_S^\pm(t)
  e^{-i\omega_S^\pm t} |2, \mathbf{p} + \hbar\mathbf{k}_P\rangle,
\end{aligned}
\end{align}
where $\hat V$ is the coupling operator in rotating wave approximation
(RWA); the Rabi frequencies
\begin{align}\label{eq:rabi-freqs}
  \Omega_P(t) = - \frac{\mathbf d_{21} \mathbf E_P}{\hbar}, \quad
  \Omega_S^\pm(t) = - \frac{\mathbf d_{23}^\pm \mathbf
    E_S^\pm}{\hbar},
\end{align}
are assumed to be real-valued; $\mathbf{d}_{21}$,
$\mathbf{d}_{23}^\pm$ are the matrix elements of the dipole moment
operator. Rabi frequencies \eqref{eq:rabi-freqs} evolve in time
together with the electric field components $\mathbf E_P$, $\mathbf
E_S^+$ and $\mathbf E_S^-$, and have Gaussian envelopes arranged in a
counterintuitive sequence as shown in Fig.~\ref{fig:tripod}(c):
\begin{align}\label{eq:pulseforms}
\begin{aligned}
  &\Omega_P(t) = \Omega_{P0} e^{ -(t-t_P)^2/ 2T_P^2 }, \\
  &\Omega_S^\pm(t) = \Omega_{S0}^\pm e^{ -(t-t_S)^2/ 2T_S^2 },
\end{aligned}
\end{align}
where $t_S < t_P$; here $2T_P$, $2T_S$ are the corresponding pulse
widths.

In addition to the atom-field coupling $\hat V$, the total Hamiltonian
for the atom-field system
\begin{align*}
  \hat H = \hat H_0 + \frac{\hat{\mathbf P}^2}{2M} + \hat V
\end{align*}
includes the kinetic term $\hat{\mathbf P}^2/(2M)$, and the energy
$\hat H_0$ of a non-moving atom with the internal state energies
$E_1$, $E_2$, $E_3^+$ and $E_3^-$. As long as spontaneous emission is
not taken into account, the atomic states in the tripod system form a
closed family of momentum $\mathbf{p}$:
\begin{multline}\label{eq:family}
  \mathcal F(\mathbf{p}) = \{ |1, \mathbf{p}\rangle, |2, \mathbf{p} +
  \hbar\mathbf{k}_P\rangle,
  \\
  |3^+, \mathbf{p} + \hbar\mathbf{k}_P - \hbar\mathbf{k}_S\rangle,
  |3^-, \mathbf{p} + \hbar\mathbf{k}_P + \hbar\mathbf{k}_S\rangle \}.
\end{multline}
As a result, in the basis of four bare states,
\begin{align}
\begin{aligned}
  |a_1 \rangle = &\exp\biggl[-i\biggl(\frac{E_1}{\hbar} +
  \frac{\mathbf{p}^2}{2M\hbar}\biggr) t\biggr] |1, \mathbf{p}\rangle,
  \\
  |a_2 \rangle = &\exp\biggl[-i\biggl(\frac{E_2}{\hbar} +
  \frac{\mathbf{p}^2}{2M\hbar} + \Delta_P \biggr) t\biggr] |2,
  \mathbf{p} + \hbar\mathbf{k}_P\rangle,
  \\
  |a_3^\pm \rangle = &\exp\biggl[-i\biggl(\frac{E_3^\pm}{\hbar} +
  \frac{\mathbf{p}^2}{2M\hbar} + \Delta_P - \Delta_S^\pm \biggr)
  t\biggr] \\
  &\times |3^\pm, \mathbf{p} + \hbar\mathbf{k}_P \mp
  \hbar\mathbf{k}_S\rangle,
\end{aligned}
\end{align}
the dynamics of the atom is described by the atomic Hamiltonian
\begin{align}\label{eq:Ham_basic}
  H = \frac{\hbar}{2} \begin{pmatrix}
    0 & \Omega_P(t) & 0 & 0
    \\
    \Omega_P(t) & -2\tilde\Delta_P &
    \Omega_S^+(t) & \Omega_S^-(t)
    \\
    0 & \Omega_S^+(t) & 2\delta_S^+ & 0
    \\
    0 & \Omega_S^-(t) & 0 & 2\delta_S^-
  \end{pmatrix},
\end{align}
with the following detunings:
\begin{align}
\begin{aligned}
  \tilde\Delta_P &= \Delta_P - \frac{k_P p_y}{M} - \omega^R_P,
  \\
  \delta_S^\pm &= \Delta_S^\pm - \Delta_P + \frac{k_P p_y \mp k_S
    p_z}{M} + \omega^R_P + \omega^R_S.
\end{aligned}
\end{align}
Here, $p_y$, $p_z$ are projections of momentum $\mathbf{p}$ on axes
$Oy$ and $Oz$, respectively; $\Delta_P = \omega_P - \omega_{21}$,
$\Delta_S^\pm = \omega_S^\pm - \omega_{23}^\pm$ are the laser
detunings; $\omega^R_P = \hbar k_P^2/(2M)$, $\omega^R_S = \hbar
k_S^2/(2M)$ are the one-photon recoil frequencies.

\section{Suppression of upper-state decay}\label{sec:no_decay}

The first step of a cooling cycle demands that the contribution of
upper-state decay is as low as possible, because the decay broadens
the velocity spread of population transfer and thus suppresses the
control required by subrecoil cooling. To avoid the undesirable
spontaneous decay, sufficiently large upper-state detunings $\Delta_P$
are commonly utilized. Here such approach means that we need to
satisfy the conditions
\begin{align}
  |\Delta_P| \gg \frac{k_Pp_y}{M}, \frac{k_Sp_z}{M}, \omega^R_P,
  \omega^R_S.
\end{align}
Then the upper state is adiabatically eliminated from the
Hamiltonian~\eqref{eq:Ham_basic} and we can write
\begin{align}\label{eq:ad_el}
  \langle a_2 |\Psi \rangle \approx \frac{\Omega_P(t)}{2\Delta_P}
  \langle a_1 |\Psi \rangle + \frac{\Omega_S^+(t)}{2\Delta_P} \langle
  a_3^+ |\Psi \rangle + \frac{\Omega_S^-(t)}{2\Delta_P} \langle a_3^-
  |\Psi \rangle,
\end{align}
where $|\Psi \rangle$ is the wave function of an atom. In turn,
Eq.~\eqref{eq:ad_el} relies on the adiabaticity constraint
\begin{align}
  |\langle a_2 | \frac{d}{dt} |\Psi \rangle | \ll |\Delta_P| |\langle
  a_2 |\Psi \rangle|,
\end{align}
which gives the necessary conditions for the validity of the
upper-state elimination, namely
\begin{align}
  |\Delta_P| \gg \Omega_{P0}, \Omega_{S0}^\pm, |\delta_S^\pm|, T^{-1}.
\end{align}
The latter term shows that the envelopes of the laser pulses should
evolve in time with a rate that is much smaller than the upper-state
detuning in frequency units, whereas the other terms in the right-hand
side respond to the splitting of the atomic levels. Then, the reduced
Hamiltonian in the basis of states $\{|1\rangle, |3^+\rangle,
|3^-\rangle\}$ is written as
\begin{align}\label{eq:Ham_red}
  \tilde H = \frac{\hbar}{2} \begin{pmatrix}
    \dfrac{\Omega_P(t)^2}{2\Delta_P} & \dfrac{\Omega_P(t)
      \Omega_S^+(t)}{2\Delta_P} & \dfrac{\Omega_P(t)
      \Omega_S^-(t)}{2\Delta_P}
    \\
    \dfrac{\Omega_P(t) \Omega_S^+(t)}{2\Delta_P} & 2\delta_S^+ +
    \dfrac{\Omega_S^+(t)^2}{2\Delta_P} & \dfrac{\Omega_S^+(t)
      \Omega_S^-(t)}{2\Delta_P}
    \\
    \dfrac{\Omega_P(t) \Omega_S^-(t)}{2\Delta_P} &
    \dfrac{\Omega_S^+(t) \Omega_S^-(t)}{2\Delta_P} & 2\delta_S^- +
    \dfrac{\Omega_S^-(t)^2}{2\Delta_P}
  \end{pmatrix}.
\end{align}

The contribution of spontaneous decay is estimated as the loss of
population from the upper state of natural width $\Gamma$ during
\mbox{STIRAP} process. With help of the density operator $\hat
\sigma$, whose matrix elements are
\begin{align}\label{eq:sigma}
  \sigma_{ij}(\mathbf{p}) = \langle a_i |\hat \sigma| a_j \rangle,
  \quad i, j = 1, 2, 3^+, 3^-,
\end{align}
the population loss is given by
\begin{align}
  \frac{d}{dt}\sigma_\mathrm{sp}(\mathbf{p}) = \Gamma
  \sigma_{22}(\mathbf{p}),
  \quad
  \sigma_\mathrm{sp}(\mathbf{p}) = \Gamma \int_0^{\Delta\tau}
  \sigma_{22}(\mathbf{p}) dt,
\end{align}
occurring during time interval $\Delta\tau$ while the \mbox{STIRAP}
pulses overlap. Because the adiabatic process takes a long time, the
overall population loss $\sigma_\mathrm{sp}(\mathbf{p})$ can not be
neglected at this point. To estimate $\sigma_\mathrm{sp}(\mathbf{p})$,
notice that the condition in Eq.~\eqref{eq:ad_el} for upper-state
elimination leads to the following inequality:
\begin{multline}
  \sigma_{22}(\mathbf{p}) \lesssim 3 \left(
    \frac{\Omega_P(t)^2}{4\Delta_P^2} \sigma_{11}(\mathbf{p}) +
    \frac{\Omega_S^+(t)^2}{4\Delta_P^2} \sigma_{33}^{++}(\mathbf{p})
  \right.
  \\
  \left. + \frac{\Omega_S^-(t)^2}{4\Delta_P^2}
    \sigma_{33}^{--}(\mathbf{p}) \right) \le 3\frac{\Omega_P(t)^2 +
    \Omega_S^+(t)^2 + \Omega_S^-(t)^2}{4\Delta_P^2}.
\end{multline}
The effect of the spontaneous decay is now estimated by
\begin{align}
  \sigma_\mathrm{sp}(\mathbf{p}) \le 3\Gamma \frac{\Omega_P(t)^2 +
    \Omega_S^+(t)^2 + \Omega_S^-(t)^2}{4\Delta_P^2} \Delta\tau,
\end{align}
and it can be neglected when $\sigma_\mathrm{sp}(\mathbf{p}) \ll 1$.
So, if the constraint
\begin{align}\label{eq:no-decay}
  \frac{|\Delta_P|}{\Gamma} \gg \frac{\Omega_P(t)^2 + \Omega_S^+(t)^2
    + \Omega_S^-(t)^2}{|\Delta_P|} \Delta\tau
\end{align}
is satisfied, then the upper-state decay can be neglected from
consideration.

\section{Elementary cooling cycle}\label{sec:res_group}

In the first cooling step, \mbox{STIRAP} only accomplishes a transfer
of atoms through the dark state formed by the ground states of the
tripod system, thereby determining the velocity selectivity of the
transfer. As a combination of the original ground state $|1\rangle$
with either the $|3^+\rangle$ or $|3^-\rangle$ state, the dark state
occurs under the condition of the two-photon resonance between
selected ground states. The associated resonant velocities follow from
the Hamiltonian in Eq.~\eqref{eq:Ham_basic} by setting
\begin{align}
  \delta_S^+ = 0 \quad \mbox{or} \quad \delta_S^- = 0.
\end{align}
The former condition corresponds to population transfer by Raman
transition $|1\rangle \leftrightarrow |3^+\rangle$, whereas the latter
corresponds to the $|1\rangle \leftrightarrow |3^-\rangle$ transition.

\begin{figure}
(a)\\ \includegraphics{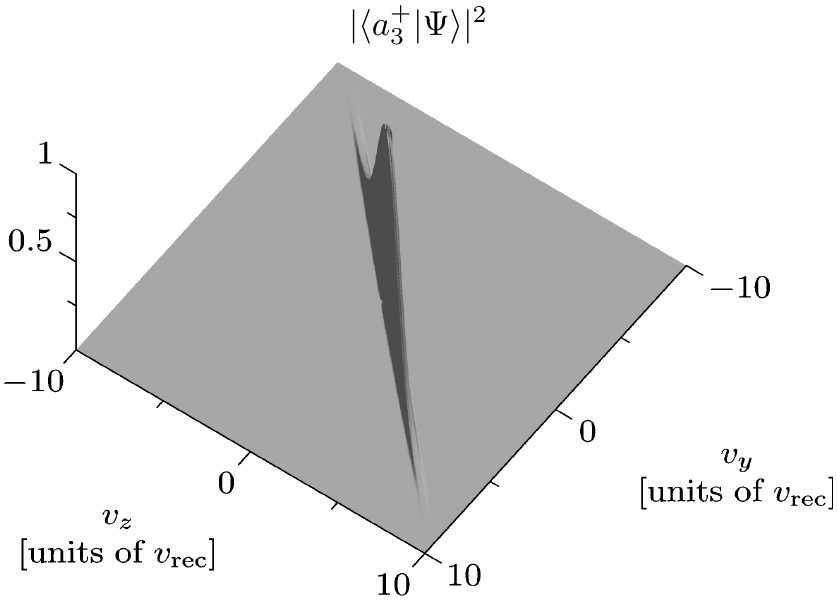}
\\
(b)\\ \includegraphics{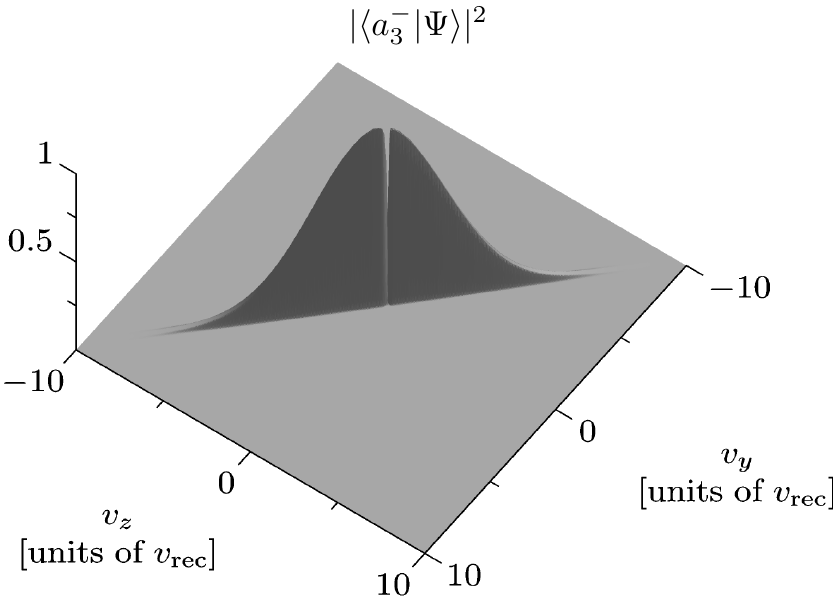}
\\
(c)\\ \includegraphics{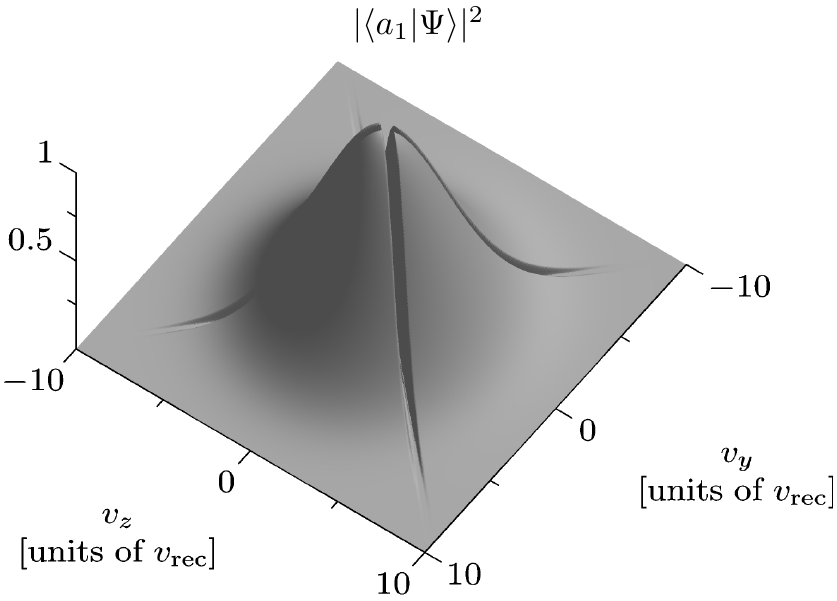}
\caption{\label{fig:transfer} (a) and (b): Parts of the initial
  velocity distribution transferred by the \mbox{STIRAP} process into
  the states $|3^+\rangle$ and $|3^-\rangle$, respectively; (c) The
  hole burning arising in state $|1\rangle$ due to these transfers.
  The scaling is given by the recoil velocity $v_\mathrm{rec} = \hbar
  k_P/M = \hbar k_S/M$ and we also set frequency $\omega^R =
  \omega^R_P = \omega^R_S$. The maximal magnitudes
  $\Omega_{P0}^2/|\Delta_P|$, $(\Omega_{S0}^\pm)^2/|\Delta_P|$ of the
  two-photon Rabi frequencies are equal to $2\omega^R$, and $\Delta_P
  - \Delta_S^\pm = \omega^R$. The \mbox{STIRAP}-pulse duration
  $T_\mathrm{pulse}$ is $96(\omega^R)^{-1}$.}
\end{figure}

Figures~\ref{fig:transfer}(a) and (b) illustrate how the atoms are
transferred into the $|3^+\rangle$ and $|3^-\rangle$ states depending
on their velocity, and the corresponding hole burning for atoms in
state $|1\rangle$ is shown in Fig.~\ref{fig:transfer}(c). The use of
both transitions for transferring the dark-state atoms from the ground
state $|1\rangle$ is an obvious advantage of the tripod system,
because atoms can be simultaneously cooled in both dimensions. Form of
the burned cross-like hole is defined by the laser configuration. The
only variable part is the position of the cross-like pattern, whose
center is given by
\begin{align}\label{eq:center}
  \delta_S^+ = \delta_S^- = 0,
\end{align}
and can be shifted by changing detunings $\Delta_P$, $\Delta_S^\pm$.

In order to return atoms from states $|3^+\rangle$ and $|3^-\rangle$
to the state $|1\rangle$, a fast optical pumping process is needed
after the \mbox{STIRAP} pulse. The $\pi$-polarized laser is switched
off, and only circularly polarized beams are left, being now tuned
into resonance. An atom of momentum $\mathbf{p}$ is excited from
either ground state $|3^+\rangle$ or $|3^-\rangle$ to the upper state
and therefore it gains a momentum kick along the $Oz$ direction.
Although the momentum kick equals $\hbar\omega_{23}/c$, one can
consider that $\omega_{23}/c \approx k_S$, and hence the atomic
momentum becomes $\mathbf{p}' = \mathbf{p} \pm \hbar \mathbf{k}_S$.
Then the atom decays into the $|1\rangle$ state emitting a spontaneous
photon of momentum $\Delta\mathbf{p}$, where $|\Delta\mathbf{p}| =
\hbar\omega_{21}/c \approx \hbar k_P$. Due to momentum conservation,
the atomic momentum changes by $-\Delta\mathbf{p}$, and the population
in state $|1\rangle$ takes the form
\begin{multline}
  \langle 1, \mathbf{p}| \hat\sigma' |1, \mathbf{p}\rangle = \langle
  1, \mathbf{p}| \hat\sigma |1, \mathbf{p}\rangle
  \\
  + \langle 3^+, \mathbf{p} - \hbar \mathbf{k}_S + \Delta\mathbf{p}'|
  \hat\sigma |3^+, \mathbf{p} - \hbar \mathbf{k}_S +
  \Delta\mathbf{p}'\rangle
  \\
  + \langle 3^-, \mathbf{p} + \hbar \mathbf{k}_S + \Delta\mathbf{p}''|
  \hat\sigma |3^-, \mathbf{p} + \hbar \mathbf{k}_S +
  \Delta\mathbf{p}''\rangle.
\end{multline}
In turn, the populations can be expressed in terms of the density
matrix elements $\sigma_{ij}(\mathbf{p})$~\eqref{eq:sigma} associated
with the momentum family $\mathcal F(\mathbf{p})$~\eqref{eq:family}.
Taking into account that states $|a_j\rangle$ ($j=1,2,3^+,3^-$) are
contained in the same family $\mathcal F(\mathbf{p})$, this leads to
the expression
\begin{multline}\label{eq:sp_decay}
  \sigma'_{11}(\mathbf{p}) = \sigma_{11}(\mathbf{p}) +
  \sigma_{33}^{++}(\mathbf{p} - \hbar \mathbf{k}_P +
  \Delta\mathbf{p}')
  \\
  + \sigma_{33}^{--}(\mathbf{p} - \hbar \mathbf{k}_P +
  \Delta\mathbf{p}''),
\end{multline}
where momentum shifts are given in relation to $\mathcal
F(\mathbf{p})$.

In contrast to the first cooling step where an atom is contained in
the same family $\mathcal F(\mathbf{p})$ during the \mbox{STIRAP}, the
optical pumping process mixes the different families $\mathcal
F(\mathbf{p})$ as shown in Eq.~\eqref{eq:sp_decay}. One can see by
averaging Eq.~\eqref{eq:sp_decay} over all possible directions of
spontaneous decay that an elementary cooling cycle generally pushes
the velocity distribution along the $Oy$ axis. If the hole burning
center $\mathbf{v}^0 = \mathbf{p}^0/M$ of atoms transferred from state
$|1\rangle$ (see Fig.~\ref{fig:transfer}(c)) is adjusted to $v_y^0 <
0$, then atoms at the left-hand wing on axis $Oy$ are pushed closer to
the zero velocity, which leads to the cooling of the atomic ensemble.
In addition, a laser configuration with a $\pi$-polarized beam in the
opposite direction of propagation and adjustment to $v_y^0 > 0$ cools
atoms also in the right-side wing of axis $Oy$. When these laser
configurations alternate, a cooling of whole the ensemble becomes
feasible.

\section{Resonant group of \mbox{STIRAP}}\label{sec:params}

The efficiency that accompanies the first step of cooling cycle relies
on both a narrow velocity range and the entire transfer of
resonant-group atoms. In fact, such entire adiabatic transfer occurs
if each laser pulse is tuned into resonance with the corresponding
atomic transition. Such an approach is efficiently applied in
Ref.~\cite{Korsunsky1996} with the aim of \mbox{VSCPT} cooling. On the
other hand, an efficient transfer of dark-state atoms takes place even
in the case of large detuning $\Delta_P$, as was successfully
demonstrated for 1D subrecoil Raman cooling by
\mbox{STIRAP}~\cite{Ivanov2012}.

For velocity-selective \mbox{STIRAP}, a transfer of dark-state atoms
from the original state $|1\rangle$ evolves with the efficiency
sensitive to the velocity of the dark state. The crossing center
defined by Eq.~\eqref{eq:center} is depleted to a greater extent,
because its velocity $\mathbf{v}^0$ is attainable for both the
$|1\rangle \leftrightarrow |3^+\rangle$ and the $|1\rangle
\leftrightarrow |3^-\rangle$ Raman transitions. For the same reason
the velocity spread of the resonant group is widest at the velocity
$\mathbf{v}^0$. Next we consider the adiabaticity criterion for the
population transfer from state $|1\rangle$ at the hole burning center,
i.e., for conditions given in Eq.~\eqref{eq:center}.

Instead of assuming the conditions in Eq.~\eqref{eq:center} directly,
we first take the more general case of $\delta_S^+ = \delta_S^-$ which
corresponds to an arbitrary velocity projection $v_y$ and
\begin{align}
  v_z = \frac{\Delta_S^+ - \Delta_S^-}{2 k_S}.
\end{align}
Further, we only consider the case of $\Delta_S^+ = \Delta_S^- =
\Delta_S$ when $v_z = 0$. Such an approach gives us a condition when
the zero-velocity atoms do not leave the $|1\rangle$ state and are
efficiently accumulated there.

To simplify the equations of motion, we get from the Hamiltonian in
Eq.~\eqref{eq:Ham_red} the relationship
\begin{multline}\label{eq:diff_pm}
  i\frac{d}{dt}\left( \Omega_S^-(t) \langle a_3^+ |\Psi \rangle -
    \Omega_S^+(t) \langle a_3^- |\Psi \rangle \right)
  \\
  = (\delta_S^+ + i\Theta(t)) \left( \Omega_S^-(t) \langle a_3^+ |\Psi
    \rangle - \Omega_S^+(t) \langle a_3^- |\Psi \rangle \right),
\end{multline}
which only requires that both Stokes pulses, $\Omega_S^+(t)$ and
$\Omega_S^-(t)$, evolve in time simultaneously:
\begin{align}\label{eq:Theta}
  \frac{\dot \Omega_S^+(t)}{\Omega_S^+(t)} = \frac{\dot
    \Omega_S^-(t)}{\Omega_S^-(t)} = \Theta(t).
\end{align}
Before the STIRAP pulse starts, the atoms are contained in state
$|1\rangle$. Hence $\langle a_3^+ |\Psi \rangle^0, \langle a_3^- |\Psi
\rangle^0 = 0$, and one obtains from Eq.~\eqref{eq:diff_pm} that
\begin{align}\label{eq:eqv-pm}
  \Omega_S^-(t) \langle a_3^+ |\Psi \rangle = \Omega_S^+(t) \langle
  a_3^- |\Psi \rangle.
\end{align}

We consider the following coupled (C) and non-coupled (NC) states of
the coupling operator in Eq.~\eqref{eq:coupling}:
\begin{align}
  & |\mathrm C \rangle = \frac{\Omega_S^+(t)}{\Omega_S(t)} |a_3^+
  \rangle + \frac{\Omega_S^-(t)}{\Omega_S(t)} |a_3^- \rangle,
  \\
  & |\mathrm{NC}\rangle = \frac{\Omega_S^-(t)}{\Omega_S(t)} |a_3^+
  \rangle - \frac{\Omega_S^+(t)}{\Omega_S(t)} |a_3^- \rangle,
\end{align}
where $\Omega_S(t)^2 = \Omega_S^+(t)^2 + \Omega_S^-(t)^2$. It follows
from Eq.~\eqref{eq:eqv-pm} that $\langle\mathrm{NC}|\Psi\rangle = 0$,
which in turn leads to relationships
\begin{align}
  \langle a_3^+|\Psi\rangle = \frac{\Omega_S^+(t)}{\Omega_S(t)}
  \langle\mathrm C |\Psi\rangle,
  \quad
  \langle a_3^-|\Psi\rangle = \frac{\Omega_S^-(t)}{\Omega_S(t)}
  \langle\mathrm C |\Psi\rangle.
\end{align}
Equation~\eqref{eq:Theta} shows that $\Omega_S^\pm(t)/\Omega_S(t)$ are
constant during the \mbox{STIRAP} process. Hence the Hamiltonian in
Eq.~\eqref{eq:Ham_red} in the basis of states $\{ |a_1\rangle,
|\mathrm C\rangle \}$ takes the form
\begin{align}\label{eq:Ham_eff}
  \hat H_\mathrm{eff}=\frac{\hbar}{2} \begin{pmatrix}
    -2\delta_0(t) & \Omega_\mathrm{eff}(t)
    \\
    \Omega_\mathrm{eff}(t) & 2(\delta_\mathrm{eff}(t) - \delta_0(t))
  \end{pmatrix}.
\end{align}
The effective detunings and the Rabi frequency are
\begin{align}
  & \delta_0(t) = -\frac{\Omega_P(t)^2}{4\Delta_P}, \quad
  \Omega_\mathrm{eff}(t) = \frac{\Omega_P(t)
    \Omega_S(t)}{2\Delta_P},\nonumber
  \\
  & \delta_\mathrm{eff}(t) = \Delta\delta + \frac{\Omega_S(t)^2 -
    \Omega_P^2(t)}{4\Delta_P},
\end{align}
where detuning $\Delta\delta$ determines the offset from the resonance
velocity:
\begin{align}\label{eq:Ddelta}
  \Delta\delta = \Delta_S - \Delta_P + \frac{k_P p_y}{M} + \omega^R_P
  + \omega^R_S = k_P (v_y - v_y^0).
\end{align}

The Hamiltonian in Eq.~\eqref{eq:Ham_eff} describes an effective
two-level system considered in Ref.~\cite{Ivanov2012}. Almost the
entire transfer of the resonant-velocity atoms occurs once the
following adiabaticity criterion is fulfilled~\cite{Ivanov2012}:
\begin{align}
  \frac{\Omega_P(t)^2 + \Omega_S(t)^2}{|\Delta_P|} \Delta\tau \gg 1.
\end{align}
The adiabaticity criterion in combination with Eq.~\eqref{eq:no-decay}
gives the conditions
\begin{align}\label{eq:ad_cr}
  \frac{|\Delta_P|}{\Gamma} \gg \frac{\Omega_P(t)^2 + \Omega_S^+(t)^2
    + \Omega_S^-(t)^2}{|\Delta_P|} \Delta\tau \gg 1.
\end{align}
This constraint gives the value of $\Delta\tau$, which is needed for
achieving an entire transfer of the resonant-group atoms. Unlike with
the normal Raman processes using square or Blackman envelopes for
pulses, the requirement of having exactly a $\pi$-pulse is not present
here. However, the appropriate \mbox{STIRAP} pulses should vary slow
enough, which makes the pulse durations larger than those of
$\pi$-pulses in normal Raman process. Hence, the \mbox{STIRAP}
transfer is suitable in cases where the cooling time is not in any
critical role, giving in exchange robustness in setting the actual
pulse durations.

The resonant-velocity group can be evaluated in terms of
$\Delta\delta$ (Eq.~\eqref{eq:Ddelta}). Taking into account that
$\Delta\delta = k_P (v_y - v_y^0)$ differs from the $\Delta\delta = 2k
(v - v_0)$ obtained for the 1D case~\cite{Ivanov2012}, the velocity
spread of the resonant group through $v_z = 0$ is given
by~\cite{Ivanov2012}
\begin{align}
\begin{aligned}
  -\frac{(\Omega_{S0}^+)^2 + (\Omega_{S0}^-)^2}{2k_P|\Delta_P|} < v_y
  - v_y^0 < \frac{\Omega_{P0}^2}{2k_P|\Delta_P|} \quad \mbox{if
    $\Delta_P > 0$},
  \\
  -\frac{\Omega_{P0}^2}{2k_P|\Delta_P|} < v_y - v_y^0 <
  \frac{(\Omega_{S0}^+)^2 + (\Omega_{S0}^-)^2}{2k_P|\Delta_P|} \quad
  \mbox{if $\Delta_P < 0$}.
\end{aligned}
\end{align}
One can see that the velocity spread defined by the two-photon Rabi
frequencies can get as narrow as needed for deep subrecoil cooling. On
the other hand, large Rabi frequencies broaden the velocity profile of
the transfer. As a result, an appropriate tuning of the pulse
intensities allows one to cool the atomic ensemble substantially below
the recoil limit.

\section{Full cooling process}\label{sec:cooling}

A single cooling cycle consists of two steps, namely the population
transfer by \mbox{STIRAP} pulse and the subsequent optical pumping
which returns the atoms back to the original internal state due to
spontaneous decay. We assume that $k_P = k_S$ and hence $\omega_P^R =
\omega_S^R = \omega^R$. The initial velocity distribution of the
atomic ensemble has the spread of $3v_\mathrm{rec}$, where
$v_\mathrm{rec}$ is the common recoil velocity of all \mbox{STIRAP}
lasers. The lasers are detuned from the upper state by $\Delta_P > 0$,
and the Rabi frequencies are given by
\begin{align}
  \Omega_{P0}^2 = (\Omega_{S0}^+)^2 = (\Omega_{S0}^-)^2 = 2 k_P
  |v_y^0| |\Delta_P|,
\end{align}
suppressing the transfer of zero-velocity atoms. The laser
configuration shown in Fig.~\ref{fig:tripod}(a) has the resonant
velocity of $v_y^0 < 0$, whereas the case of the alternated
$\pi$-polarized beam corresponds to $v_y^0 > 0$. After each sequence
of five cooling cycles with
\begin{align}\label{eq:v_set}
  |v_y^0| = 2^{k-1} v_\mathrm{rec}, \quad k = 0,\ldots,4,
\end{align}
the direction of $\pi$-polarized laser is alternated. The broad
velocity profiles in the set defined in Eq.~\eqref{eq:v_set} involve
all velocity-distributed atoms in the cooling process, whereas those
in a narrow velocity group lead to the actual deep cooling below the
recoil limit.

\begin{figure}
(a)\\ \includegraphics{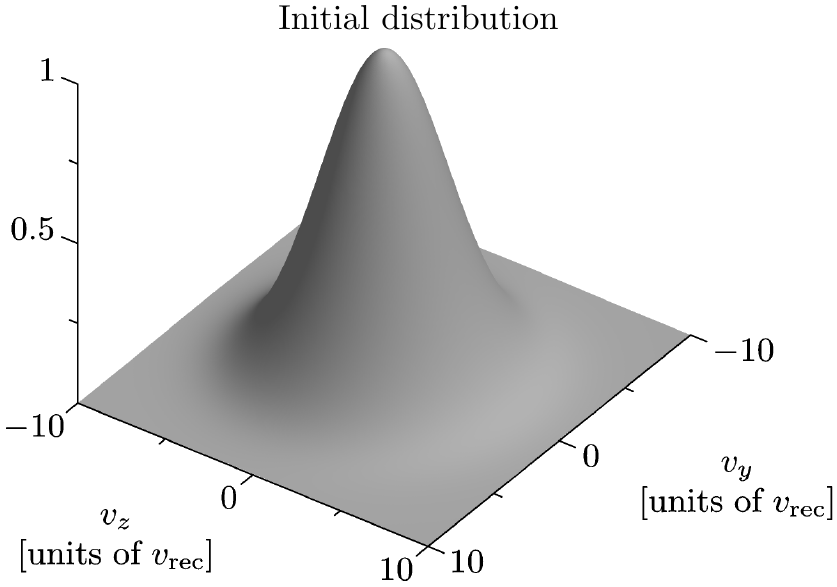}
\\
(b)\\ \includegraphics{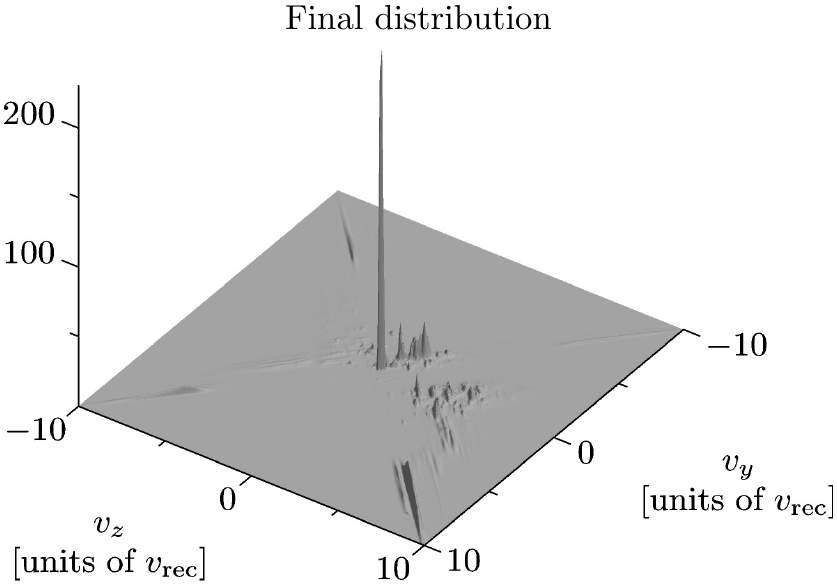}
\caption{\label{distr} The velocity distribution of an atomic ensemble
  before (a) and after (b) the 2D cooling is applied. The velocity
  spread of (a) $3v_\mathrm{rec}$ has been reduced to (b)
  $0.1v_\mathrm{rec}$ after realizing 500 single cooling cycles,
  corresponding to the effective temperature going down to
  $0.01T_\text{rec}$. The height of the central peak has increased to
  about 230 times that of the initial distribution.}
\end{figure}

The pump and the Stokes pulses have the same pulse shape with the
pulse half-widths (see Eq.~\eqref{eq:pulseforms})
\begin{align*}
  T_P = T_S = 0.5 (t_P - t_S).
\end{align*}
The duration $T_\mathrm{pulse}$ of the \mbox{STIRAP} pulses is defined
by the start and the end times
\begin{align*}
  t_\mathrm{start} = t_S - (t_P - t_S), \quad t_\mathrm{end} = t_P +
  (t_P - t_S)
\end{align*}
being equal to $3(t_P - t_S)$. As $|v_y^0|$ decreases, the magnitudes
$\Omega_{P0}^2/|\Delta_P|$ and $(\Omega_{S0}^\pm)^2/|\Delta_P|$ will
decrease as well, leading to a corresponding increase in $\Delta\tau$,
so that the adiabaticity criterion in Eq.~\eqref{eq:ad_cr} is
fulfilled. As a result, the pulse durations according to the set in
Eq.~\eqref{eq:v_set} are given by
\begin{align*}
  T_\mathrm{pulse} = 6 \cdot 2^k \tau^R \quad k = 0, \ldots, 4,
\end{align*}
where $\tau^R = (\omega^R)^{-1}$ is the recoil time.

The \mbox{STIRAP} cooling process collects atoms into a narrow peak
near the zero velocity, and the height of this peak grows
simultaneously with the number $N$ of cooling cycles. The result after
applying $N = 500$ elementary cycles of 2D \mbox{STIRAP} cooling is
shown in Fig.~\ref{distr}, where the height of the peak has become
about 230 times higher than that of the initial broad distribution.
The velocity spread of the atomic ensemble, given by
$\sigma=(\mbox{FWHM})/\sqrt{8\ln 2}$, has been reduced from
$3v_\mathrm{rec}$ to $0.1v_\mathrm{rec}$. The corresponding effective
temperature $T_\mathrm{eff}$ goes down to $0.01T_\mathrm{rec}$, where
$T_\mathrm{rec}$ is the recoil-limit temperature.

Let us consider the result of this cooling as applied to metastable Ne
atoms under the conditions in the experiment described in
Ref.~\cite{Theuer1999}. Both the $\sigma^+$- and $\sigma^-$-polarized
waves are provided by laser light at wavelength $\lambda_S=588 \text{
  nm}$, whereas laser light at $\lambda_P=617\text{ nm}$ generates the
$\pi$-polarized beam. As a result, the final effective temperature of
cooled atoms is given by
\begin{align*}
  T_\text{eff}\approx 0.01\frac{\hbar^2}{2 k_B M}(k_P^2 + k_S^2) =
  26\text{ nK},
\end{align*}
where $k_B$ is the Boltzmann constant, $M$ is the Ne atomic mass. The
duration of only \mbox{STIRAP} pulses in our scheme is 18600~$\tau^R$,
hence full cooling takes about 0.1 s.

\section{Conclusions}\label{sec:conclusion}

We have considered a variant of optical cooling based on
velocity-selective \mbox{STIRAP} transfer in a four-level tripod
system. This approach extends into 2D the recently proposed 1D cooling
method~\cite{Ivanov2012}, providing strong transverse cooling below
the recoil limit. In contrast to the normal 2D Raman
cooling~\cite{Ivanov2011}, the method is robust and versatile as long
as the adiabaticity criterion is satisfied. Strong and efficient
cooling is especially attainable at the limit of large detuning from
the upper state in the tripod configuration. The numerical results
demonstrate a 2D cooling down to $0.01T_\text{rec}$.

As a topic of discussion, we note that the success of evaporative
cooling in reaching the atomic phase-space density that is required
for quantum degeneracy has diminished strongly the original interest
in light-based cooling methods, although the increasing variety of
experiments with neutral atoms under differing circumstances is
reviving this interest. Similarly, the dynamics of light-assisted cold
atomic collisions~\cite{Weiner1999,Suominen1996b} have not been fully
explored, and in fact the issue of their character is still unresolved
experimentally~\cite{Mastwijk1998,Glover2011}. The dynamical viewpoint
involving level
crossings~\cite{Suominen1994,Suominen1996a,Suominen1998} and the
complementary view of steady-state
description~\cite{Gallagher1989,Julienne1991} have both their
supporters, and it would be of interest to examine the dependence of
the collisional atomic kinetic energy gain as a function of laser
intensity and especially detuning~\cite{Holland1994a,Holland1994b}. At
ultralow temperatures the collisions take a very different nature
compared to the more semiclassical idea of atoms approaching each
other~\cite{Weiner1999,Burnett1996}; also the interesting question
about the role of the higher partial waves tends to disappear when
quantum statistics steps in, and energies limit the processes to the
$s$-wave only~\cite{Piilo2006}. Such studies are an example where one
could apply such light-based cooling methods as we have proposed. A
special feature in the STIRAP-based cooling is the possibility to use
large detunings. This reduces the role of light-assisted collisions or
reabsorption of scattered photons in the cooling process, allowing
higher densities than available at standard magneto-optical traps,
while collisions and other properties of a cold but still
non-degenerate gas can be analysed with separate probe lasers.

\section{Acknowledgments}

This research was supported by the Finnish Academy of Science and
Letters, CIMO, and the Academy of Finland, grant 133682.

\bibliography{STIRAP-cooling}

%Merlin.mbs v4.21 2009-07-09.
\begin{thebibliography}{10}%
\makeatletter
\providecommand \@ifxundefined [1]{%
 \ifx #1\undefined \expandafter \@firstoftwo
 \else \expandafter \@secondoftwo
\fi
}%
\providecommand \@ifnum [1]{%
 \ifnum #1\expandafter \@firstoftwo
 \else \expandafter \@secondoftwo
\fi
}%
\providecommand \enquote [1]{``#1''}%
\providecommand \bibnamefont  [1]{#1}%
\providecommand \bibfnamefont [1]{#1}%
\providecommand \citenamefont [1]{#1}%
\providecommand\href[0]{\@sanitize\@href}%
\providecommand\@href[1]{\endgroup\@@startlink{#1}\endgroup\@@href}%
\providecommand\@@href[1]{#1\@@endlink}%
\providecommand \@sanitize [0]{\begingroup\catcode`\&12\catcode`\#12\relax}%
\@ifxundefined \pdfoutput {\@firstoftwo}{%
 \@ifnum{\z@=\pdfoutput}{\@firstoftwo}{\@secondoftwo}%
}{%
 \providecommand\@@startlink[1]{\leavevmode\special{html:<a href="#1">}}%
 \providecommand\@@endlink[0]{\special{html:</a>}}%
}{%
 \providecommand\@@startlink[1]{%
  \leavevmode
  \pdfstartlink
   attr{/Border[0 0 1 ]/H/I/C[0 1 1]}%
   user{/Subtype/Link/A<</Type/Action/S/URI/URI(#1)>>}%
  \relax
 }%
 \providecommand\@@endlink[0]{\pdfendlink}%
}%
\providecommand \url  [0]{\begingroup\@sanitize \@url }%
\providecommand \@url [1]{\endgroup\@href {#1}{\urlprefix}}%
\providecommand \urlprefix [0]{URL }%
\providecommand \Eprint[0]{\href }%
\@ifxundefined \urlstyle {%
  \providecommand \doi [1]{doi:\discretionary{}{}{}#1}%
}{%
  \providecommand \doi [0]{doi:\discretionary{}{}{}\begingroup
  \urlstyle{rm}\Url }%
}%
\providecommand \doibase [0]{http://dx.doi.org/}%
\providecommand \Doi[1]{\href{\doibase#1}}%
\providecommand \bibAnnote [3]{%
  \BibitemShut{#1}%
  \begin{quotation}\noindent
    \textsc{Key:}\ #2\\\textsc{Annotation:}\ #3%
  \end{quotation}%
}%
\providecommand \bibAnnoteFile [2]{%
  \IfFileExists{#2}{\bibAnnote {#1} {#2} {\input{#2}}}{}%
}%
\providecommand \typeout [0]{\immediate \write \m@ne }%
\providecommand \selectlanguage [0]{\@gobble}%
\providecommand \bibinfo [0]{\@secondoftwo}%
\providecommand \bibfield [0]{\@secondoftwo}%
\providecommand \translation [1]{[#1]}%
\providecommand \BibitemOpen[0]{}%
\providecommand \bibitemStop [0]{}%
\providecommand \bibitemNoStop [0]{.\EOS\space}%
\providecommand \EOS [0]{\spacefactor3000\relax}%
\providecommand \BibitemShut [1]{\csname bibitem#1\endcsname}%
%</preamble>
\bibitem{Weiner1999}%
  \BibitemOpen
  \bibfield{author}{%
  \bibinfo {author} {\bibfnamefont{J.}~\bibnamefont{Weiner}}, \bibinfo {author}
  {\bibfnamefont{V.~S.}\ \bibnamefont{Bagnato}}, \bibinfo {author}
  {\bibfnamefont{S.}~\bibnamefont{Zilio}},\ and\ \bibinfo {author}
  {\bibfnamefont{P.~S.}\ \bibnamefont{Julienne}},\ }%
  \bibfield{journal}{%
  \Doi{10.1103/RevModPhys.71.1}{\bibinfo {journal} {Rev. Mod. Phys.}}\ }%
  \textbf{\bibinfo {volume} {71}},\ \bibinfo {pages} {1} (\bibinfo {year}
  {1999})%
  \bibAnnoteFile{NoStop}{Weiner1999}%
\bibitem{Calsamiglia2001}%
  \BibitemOpen
  \bibfield{author}{%
  \bibinfo {author} {\bibfnamefont{J.}~\bibnamefont{Calsamiglia}}, \bibinfo
  {author} {\bibfnamefont{M.}~\bibnamefont{Mackie}},\ and\ \bibinfo {author}
  {\bibfnamefont{K.-A.}\ \bibnamefont{Suominen}},\ }%
  \bibfield{journal}{%
  \Doi{10.1103/PhysRevLett.87.160403}{\bibinfo {journal} {Phys. Rev. Lett.}}\
  }%
  \textbf{\bibinfo {volume} {87}},\ \bibinfo {pages} {160403} (\bibinfo {year}
  {2001})%
  \bibAnnoteFile{NoStop}{Calsamiglia2001}%
\bibitem{Jones2006}%
  \BibitemOpen
  \bibfield{author}{%
  \bibinfo {author} {\bibfnamefont{K.~M.}\ \bibnamefont{Jones}}, \bibinfo
  {author} {\bibfnamefont{E.}~\bibnamefont{Tiesinga}}, \bibinfo {author}
  {\bibfnamefont{P.~D.}\ \bibnamefont{Lett}},\ and\ \bibinfo {author}
  {\bibfnamefont{P.~S.}\ \bibnamefont{Julienne}},\ }%
  \bibfield{journal}{%
  \Doi{10.1103/RevModPhys.78.483}{\bibinfo {journal} {Rev. Mod. Phys.}}\ }%
  \textbf{\bibinfo {volume} {78}},\ \bibinfo {pages} {483} (\bibinfo {month}
  {May}\ \bibinfo {year} {2006})%
  \bibAnnoteFile{NoStop}{Jones2006}%
\bibitem{Bloch2008}%
  \BibitemOpen
  \bibfield{author}{%
  \bibinfo {author} {\bibfnamefont{I.}~\bibnamefont{Bloch}}, \bibinfo {author}
  {\bibfnamefont{J.}~\bibnamefont{Dalibard}},\ and\ \bibinfo {author}
  {\bibfnamefont{W.}~\bibnamefont{Zwerger}},\ }%
  \bibfield{journal}{%
  \Doi{10.1103/RevModPhys.80.885}{\bibinfo {journal} {Rev. Mod. Phys.}}\ }%
  \textbf{\bibinfo {volume} {80}},\ \bibinfo {pages} {885} (\bibinfo {year}
  {2008})%
  \bibAnnoteFile{NoStop}{Bloch2008}%
\bibitem{Dziarmaga2010}%
  \BibitemOpen
  \bibfield{author}{%
  \bibinfo {author} {\bibfnamefont{J.}~\bibnamefont{Dziarmaga}},\ }%
  \bibfield{journal}{%
  \Doi{10.1080/00018732.2010.514702}{\bibinfo {journal} {Adv. Phys.}}\ }%
  \textbf{\bibinfo {volume} {59}},\ \bibinfo {pages} {1063} (\bibinfo {year}
  {2010})%
  \bibAnnoteFile{NoStop}{Dziarmaga2010}%
\bibitem{Nielsen2000}%
  \BibitemOpen
  \bibfield{author}{%
  \bibinfo {author} {\bibfnamefont{M.~A.}\ \bibnamefont{Nielsen}}\ and\
  \bibinfo {author} {\bibfnamefont{I.~L.}\ \bibnamefont{Chuang}},\ }%
  \emph{\bibinfo {title} {Quantum Computation and Quantum Information}}\
  (\bibinfo {publisher} {Cambridge University Press},\ \bibinfo {year} {2000})%
  \bibAnnoteFile{NoStop}{Nielsen2000}%
\bibitem{Stenholm2005}%
  \BibitemOpen
  \bibfield{author}{%
  \bibinfo {author} {\bibfnamefont{S.}~\bibnamefont{Stenholm}}\ and\ \bibinfo
  {author} {\bibfnamefont{K.-A.}\ \bibnamefont{Suominen}},\ }%
  \emph{\bibinfo {title} {Quantum Approach to Informatics}}\ (\bibinfo
  {publisher} {John Wiley \& Sons},\ \bibinfo {year} {2005})%
  \bibAnnoteFile{NoStop}{Stenholm2005}%
\bibitem{Audoin2001}%
  \BibitemOpen
  \bibfield{author}{%
  \bibinfo {author} {\bibfnamefont{C.}~\bibnamefont{Audoin}}\ and\ \bibinfo
  {author} {\bibfnamefont{B.}~\bibnamefont{Guinot}},\ }%
  \emph{\bibinfo {title} {The Measurement of Time: Time, Frequency, and the
  Atomic Clock}}\ (\bibinfo {publisher} {Cambridge University Press},\ \bibinfo
  {year} {2001})%
  \bibAnnoteFile{NoStop}{Audoin2001}%
\bibitem{Ye2008}%
  \BibitemOpen
  \bibfield{author}{%
  \bibinfo {author} {\bibfnamefont{J.}~\bibnamefont{{Ye}}}, \bibinfo {author}
  {\bibfnamefont{H.~J.}\ \bibnamefont{{Kimble}}},\ and\ \bibinfo {author}
  {\bibfnamefont{H.}~\bibnamefont{{Katori}}},\ }%
  \bibfield{journal}{%
  \Doi{10.1126/science.1148259}{\bibinfo {journal} {Science}}\ }%
  \textbf{\bibinfo {volume} {320}},\ \bibinfo {pages} {1734} (\bibinfo {year}
  {2008})%
  \bibAnnoteFile{NoStop}{Ye2008}%
\bibitem{Derevianko2011}%
  \BibitemOpen
  \bibfield{author}{%
  \bibinfo {author} {\bibfnamefont{A.}~\bibnamefont{Derevianko}}\ and\ \bibinfo
  {author} {\bibfnamefont{H.}~\bibnamefont{Katori}},\ }%
  \bibfield{journal}{%
  \Doi{10.1103/RevModPhys.83.331}{\bibinfo {journal} {Rev. Mod. Phys.}}\ }%
  \textbf{\bibinfo {volume} {83}},\ \bibinfo {pages} {331} (\bibinfo {year}
  {2011})%
  \bibAnnoteFile{NoStop}{Derevianko2011}%
\bibitem{Cohen2011}%
  \BibitemOpen
  \bibfield{author}{%
  \bibinfo {author} {\bibfnamefont{C.}~\bibnamefont{Cohen-Tannoudji}}\ and\
  \bibinfo {author} {\bibfnamefont{D.}~\bibnamefont{Gu{\'e}ry-Odelin}},\ }%
  \emph{\bibinfo {title} {Advances In Atomic Physics: An Overview}}\ (\bibinfo
  {publisher} {World Scientific},\ \bibinfo {year} {2011})%
  \bibAnnoteFile{NoStop}{Cohen2011}%
\bibitem{Johnson1998}%
  \BibitemOpen
  \bibfield{author}{%
  \bibinfo {author} {\bibfnamefont{K.~S.}\ \bibnamefont{Johnson}}, \bibinfo
  {author} {\bibfnamefont{J.~H.}\ \bibnamefont{Thywissen}}, \bibinfo {author}
  {\bibfnamefont{N.~H.}\ \bibnamefont{Dekker}}, \bibinfo {author}
  {\bibfnamefont{K.~K.}\ \bibnamefont{Berggren}}, \bibinfo {author}
  {\bibfnamefont{A.~P.}\ \bibnamefont{Chu}}, \bibinfo {author}
  {\bibfnamefont{R.}~\bibnamefont{Younkin}},\ and\ \bibinfo {author}
  {\bibfnamefont{M.}~\bibnamefont{Prentiss}},\ }%
  \bibfield{journal}{%
  \Doi{10.1126/science.280.5369.1583}{\bibinfo {journal} {Science}}\ }%
  \textbf{\bibinfo {volume} {280}},\ \bibinfo {pages} {1583} (\bibinfo {year}
  {1998})%
  \bibAnnoteFile{NoStop}{Johnson1998}%
\bibitem{Theuer1999}%
  \BibitemOpen
  \bibfield{author}{%
  \bibinfo {author} {\bibfnamefont{H.}~\bibnamefont{Theuer}}, \bibinfo {author}
  {\bibfnamefont{R.}~\bibnamefont{Unanyan}}, \bibinfo {author}
  {\bibfnamefont{C.}~\bibnamefont{Habscheid}}, \bibinfo {author}
  {\bibfnamefont{K.}~\bibnamefont{Klein}},\ and\ \bibinfo {author}
  {\bibfnamefont{K.}~\bibnamefont{Bergmann}},\ }%
  \bibfield{journal}{%
  \Doi{10.1364/OE.4.000077}{\bibinfo {journal} {Opt. Express}}\ }%
  \textbf{\bibinfo {volume} {4}},\ \bibinfo {pages} {77} (\bibinfo {year}
  {1999})%
  \bibAnnoteFile{NoStop}{Theuer1999}%
\bibitem{Partlow2004}%
  \BibitemOpen
  \bibfield{author}{%
  \bibinfo {author} {\bibfnamefont{M.}~\bibnamefont{Partlow}}, \bibinfo
  {author} {\bibfnamefont{X.}~\bibnamefont{Miao}}, \bibinfo {author}
  {\bibfnamefont{J.}~\bibnamefont{Bochmann}}, \bibinfo {author}
  {\bibfnamefont{M.}~\bibnamefont{Cashen}},\ and\ \bibinfo {author}
  {\bibfnamefont{H.}~\bibnamefont{Metcalf}},\ }%
  \bibfield{journal}{%
  \Doi{10.1103/PhysRevLett.93.213004}{\bibinfo {journal} {Phys. Rev. Lett.}}\
  }%
  \textbf{\bibinfo {volume} {93}},\ \bibinfo {pages} {213004} (\bibinfo {year}
  {2004})%
  \bibAnnoteFile{NoStop}{Partlow2004}%
\bibitem{Lewenstein2012}%
  \BibitemOpen
  \bibfield{author}{%
  \bibinfo {author} {\bibfnamefont{M.}~\bibnamefont{Lewenstein}}, \bibinfo
  {author} {\bibfnamefont{A.}~\bibnamefont{Sanpera}},\ and\ \bibinfo {author}
  {\bibfnamefont{V.}~\bibnamefont{Ahufinger}},\ }%
  \emph{\bibinfo {title} {Ultracold Atoms in Optical Lattices}}\ (\bibinfo
  {publisher} {Oxford University Press},\ \bibinfo {year} {2012})%
  \bibAnnoteFile{NoStop}{Lewenstein2012}%
\bibitem{Morigi1997}%
  \BibitemOpen
  \bibfield{author}{%
  \bibinfo {author} {\bibfnamefont{G.}~\bibnamefont{Morigi}}, \bibinfo {author}
  {\bibfnamefont{J.~I.}\ \bibnamefont{Cirac}}, \bibinfo {author}
  {\bibfnamefont{M.}~\bibnamefont{Lewenstein}},\ and\ \bibinfo {author}
  {\bibfnamefont{P.}~\bibnamefont{Zoller}},\ }%
  \bibfield{journal}{%
  \Doi{10.1209/epl/i1997-00306-3}{\bibinfo {journal} {Europhys. Lett.}}\ }%
  \textbf{\bibinfo {volume} {39}},\ \bibinfo {pages} {13} (\bibinfo {year}
  {1997})%
  \bibAnnoteFile{NoStop}{Morigi1997}%
\bibitem{Schmidt-Kaler2001}%
  \BibitemOpen
  \bibfield{author}{%
  \bibinfo {author} {\bibfnamefont{F.}~\bibnamefont{Schmidt-Kaler}}, \bibinfo
  {author} {\bibfnamefont{J.}~\bibnamefont{Eschner}}, \bibinfo {author}
  {\bibfnamefont{G.}~\bibnamefont{Morigi}}, \bibinfo {author}
  {\bibfnamefont{C.~F.}\ \bibnamefont{Roos}}, \bibinfo {author}
  {\bibfnamefont{D.}~\bibnamefont{Leibfried}}, \bibinfo {author}
  {\bibfnamefont{A.}~\bibnamefont{Mundt}},\ and\ \bibinfo {author}
  {\bibfnamefont{R.}~\bibnamefont{Blatt}},\ }%
  \bibfield{journal}{%
  \Doi{10.1007/s003400100721}{\bibinfo {journal} {Appl. Phys. B}}\ }%
  \textbf{\bibinfo {volume} {73}},\ \bibinfo {pages} {807} (\bibinfo {year}
  {2001})%
  \bibAnnoteFile{NoStop}{Schmidt-Kaler2001}%
\bibitem{Leibfried2003}%
  \BibitemOpen
  \bibfield{author}{%
  \bibinfo {author} {\bibfnamefont{D.}~\bibnamefont{Leibfried}}, \bibinfo
  {author} {\bibfnamefont{R.}~\bibnamefont{Blatt}}, \bibinfo {author}
  {\bibfnamefont{C.}~\bibnamefont{Monroe}},\ and\ \bibinfo {author}
  {\bibfnamefont{D.}~\bibnamefont{Wineland}},\ }%
  \bibfield{journal}{%
  \Doi{10.1103/RevModPhys.75.281}{\bibinfo {journal} {Rev. Mod. Phys.}}\ }%
  \textbf{\bibinfo {volume} {75}},\ \bibinfo {pages} {281} (\bibinfo {year}
  {2003})%
  \bibAnnoteFile{NoStop}{Leibfried2003}%
\bibitem{Cirac1995}%
  \BibitemOpen
  \bibfield{author}{%
  \bibinfo {author} {\bibfnamefont{J.~I.}\ \bibnamefont{Cirac}}\ and\ \bibinfo
  {author} {\bibfnamefont{P.}~\bibnamefont{Zoller}},\ }%
  \bibfield{journal}{%
  \Doi{10.1103/PhysRevLett.74.4091}{\bibinfo {journal} {Phys. Rev. Lett.}}\ }%
  \textbf{\bibinfo {volume} {74}},\ \bibinfo {pages} {4091} (\bibinfo {year}
  {1995})%
  \bibAnnoteFile{NoStop}{Cirac1995}%
\bibitem{Haffner2008}%
  \BibitemOpen
  \bibfield{author}{%
  \bibinfo {author} {\bibfnamefont{H.}~\bibnamefont{H{\"a}ffner}}, \bibinfo
  {author} {\bibfnamefont{C.}~\bibnamefont{Roos}},\ and\ \bibinfo {author}
  {\bibfnamefont{R.}~\bibnamefont{Blatt}},\ }%
  \bibfield{journal}{%
  \Doi{10.1016/j.physrep.2008.09.003}{\bibinfo {journal} {Phys. Rep.}}\ }%
  \textbf{\bibinfo {volume} {469}},\ \bibinfo {pages} {155 } (\bibinfo {year}
  {2008})%
  \bibAnnoteFile{NoStop}{Haffner2008}%
\bibitem{Roghani2008}%
  \BibitemOpen
  \bibfield{author}{%
  \bibinfo {author} {\bibfnamefont{M.}~\bibnamefont{Roghani}}\ and\ \bibinfo
  {author} {\bibfnamefont{H.}~\bibnamefont{Helm}},\ }%
  \bibfield{journal}{%
  \Doi{10.1103/PhysRevA.77.043418}{\bibinfo {journal} {Phys. Rev. A}}\ }%
  \textbf{\bibinfo {volume} {77}},\ \bibinfo {pages} {043418} (\bibinfo {year}
  {2008})%
  \bibAnnoteFile{NoStop}{Roghani2008}%
\bibitem{Roghani2011}%
  \BibitemOpen
  \bibfield{author}{%
  \bibinfo {author} {\bibfnamefont{M.}~\bibnamefont{Roghani}}, \bibinfo
  {author} {\bibfnamefont{H.}~\bibnamefont{Helm}},\ and\ \bibinfo {author}
  {\bibfnamefont{H.-P.}\ \bibnamefont{Breuer}},\ }%
  \bibfield{journal}{%
  \Doi{10.1103/PhysRevLett.106.040502}{\bibinfo {journal} {Phys. Rev. Lett.}}\
  }%
  \textbf{\bibinfo {volume} {106}},\ \bibinfo {pages} {040502} (\bibinfo {year}
  {2011})%
  \bibAnnoteFile{NoStop}{Roghani2011}%
\bibitem{Roghani2012}%
  \BibitemOpen
  \bibfield{author}{%
  \bibinfo {author} {\bibfnamefont{M.}~\bibnamefont{Roghani}}, \bibinfo
  {author} {\bibfnamefont{H.-P.}\ \bibnamefont{Breuer}},\ and\ \bibinfo
  {author} {\bibfnamefont{H.}~\bibnamefont{Helm}},\ }%
  \bibfield{journal}{%
  \Doi{10.1103/PhysRevA.85.012313}{\bibinfo {journal} {Phys. Rev. A}}\ }%
  \textbf{\bibinfo {volume} {85}},\ \bibinfo {pages} {012313} (\bibinfo {year}
  {2012})%
  \bibAnnoteFile{NoStop}{Roghani2012}%
\bibitem{Ivanov2011}%
  \BibitemOpen
  \bibfield{author}{%
  \bibinfo {author} {\bibfnamefont{V.~S.}\ \bibnamefont{Ivanov}}, \bibinfo
  {author} {\bibfnamefont{{\relax Yu}.~V.}\ \bibnamefont{Rozhdestvensky}},\
  and\ \bibinfo {author} {\bibfnamefont{K.-A.}\ \bibnamefont{Suominen}},\ }%
  \bibfield{journal}{%
  \Doi{10.1103/PhysRevA.83.023407}{\bibinfo {journal} {Phys. Rev. A}}\ }%
  \textbf{\bibinfo {volume} {83}},\ \bibinfo {pages} {023407} (\bibinfo {year}
  {2011})%
  \bibAnnoteFile{NoStop}{Ivanov2011}%
\bibitem{Ivanov2012}%
  \BibitemOpen
  \bibfield{author}{%
  \bibinfo {author} {\bibfnamefont{V.~S.}\ \bibnamefont{Ivanov}}, \bibinfo
  {author} {\bibfnamefont{{\relax Yu}.~V.}\ \bibnamefont{Rozhdestvensky}},\
  and\ \bibinfo {author} {\bibfnamefont{K.-A.}\ \bibnamefont{Suominen}},\ }%
  \bibfield{journal}{%
  \Doi{10.1103/PhysRevA.85.033422}{\bibinfo {journal} {Phys. Rev. A}}\ }%
  \textbf{\bibinfo {volume} {85}},\ \bibinfo {pages} {033422} (\bibinfo {year}
  {2012})%
  \bibAnnoteFile{NoStop}{Ivanov2012}%
\bibitem{Aspect1988}%
  \BibitemOpen
  \bibfield{author}{%
  \bibinfo {author} {\bibfnamefont{A.}~\bibnamefont{Aspect}}, \bibinfo {author}
  {\bibfnamefont{E.}~\bibnamefont{Arimondo}}, \bibinfo {author}
  {\bibfnamefont{R.}~\bibnamefont{Kaiser}}, \bibinfo {author}
  {\bibfnamefont{N.}~\bibnamefont{Vansteenkiste}},\ and\ \bibinfo {author}
  {\bibfnamefont{C.}~\bibnamefont{Cohen-Tannoudji}},\ }%
  \bibfield{journal}{%
  \Doi{10.1103/PhysRevLett.61.826}{\bibinfo {journal} {Phys. Rev. Lett.}}\ }%
  \textbf{\bibinfo {volume} {61}},\ \bibinfo {pages} {826} (\bibinfo {year}
  {1988})%
  \bibAnnoteFile{NoStop}{Aspect1988}%
\bibitem{Kasevich1992}%
  \BibitemOpen
  \bibfield{author}{%
  \bibinfo {author} {\bibfnamefont{M.}~\bibnamefont{Kasevich}}\ and\ \bibinfo
  {author} {\bibfnamefont{S.}~\bibnamefont{Chu}},\ }%
  \bibfield{journal}{%
  \Doi{10.1103/PhysRevLett.69.1741}{\bibinfo {journal} {Phys. Rev. Lett.}}\ }%
  \textbf{\bibinfo {volume} {69}},\ \bibinfo {pages} {1741} (\bibinfo {year}
  {1992})%
  \bibAnnoteFile{NoStop}{Kasevich1992}%
\bibitem{Davidson1994}%
  \BibitemOpen
  \bibfield{author}{%
  \bibinfo {author} {\bibfnamefont{N.}~\bibnamefont{Davidson}}, \bibinfo
  {author} {\bibfnamefont{H.~J.}\ \bibnamefont{Lee}}, \bibinfo {author}
  {\bibfnamefont{M.}~\bibnamefont{Kasevich}},\ and\ \bibinfo {author}
  {\bibfnamefont{S.}~\bibnamefont{Chu}},\ }%
  \bibfield{journal}{%
  \Doi{10.1103/PhysRevLett.72.3158}{\bibinfo {journal} {Phys. Rev. Lett.}}\ }%
  \textbf{\bibinfo {volume} {72}},\ \bibinfo {pages} {3158} (\bibinfo {year}
  {1994})%
  \bibAnnoteFile{NoStop}{Davidson1994}%
\bibitem{Boyer2004}%
  \BibitemOpen
  \bibfield{author}{%
  \bibinfo {author} {\bibfnamefont{V.}~\bibnamefont{Boyer}}, \bibinfo {author}
  {\bibfnamefont{L.~J.}\ \bibnamefont{Lising}}, \bibinfo {author}
  {\bibfnamefont{S.~L.}\ \bibnamefont{Rolston}},\ and\ \bibinfo {author}
  {\bibfnamefont{W.~D.}\ \bibnamefont{Phillips}},\ }%
  \bibfield{journal}{%
  \Doi{10.1103/PhysRevA.70.043405}{\bibinfo {journal} {Phys. Rev. A}}\ }%
  \textbf{\bibinfo {volume} {70}},\ \bibinfo {pages} {043405} (\bibinfo {year}
  {2004})%
  \bibAnnoteFile{NoStop}{Boyer2004}%
\bibitem{Machholm2001}%
  \BibitemOpen
  \bibfield{author}{%
  \bibinfo {author} {\bibfnamefont{M.}~\bibnamefont{Machholm}}, \bibinfo
  {author} {\bibfnamefont{P.~S.}\ \bibnamefont{Julienne}},\ and\ \bibinfo
  {author} {\bibfnamefont{K.-A.}\ \bibnamefont{Suominen}},\ }%
  \bibfield{journal}{%
  \Doi{10.1103/PhysRevA.64.033425}{\bibinfo {journal} {Phys. Rev. A}}\ }%
  \textbf{\bibinfo {volume} {64}},\ \bibinfo {pages} {033425} (\bibinfo {year}
  {2001})%
  \bibAnnoteFile{NoStop}{Machholm2001}%
\bibitem{Machholm2002}%
  \BibitemOpen
  \bibfield{author}{%
  \bibinfo {author} {\bibfnamefont{M.}~\bibnamefont{Machholm}}, \bibinfo
  {author} {\bibfnamefont{P.~S.}\ \bibnamefont{Julienne}},\ and\ \bibinfo
  {author} {\bibfnamefont{K.-A.}\ \bibnamefont{Suominen}},\ }%
  \bibfield{journal}{%
  \Doi{10.1103/PhysRevA.65.023401}{\bibinfo {journal} {Phys. Rev. A}}\ }%
  \textbf{\bibinfo {volume} {65}},\ \bibinfo {pages} {023401} (\bibinfo {year}
  {2002})%
  \bibAnnoteFile{NoStop}{Machholm2002}%
\bibitem{Korsunsky1996}%
  \BibitemOpen
  \bibfield{author}{%
  \bibinfo {author} {\bibfnamefont{E.}~\bibnamefont{Korsunsky}},\ }%
  \bibfield{journal}{%
  \Doi{10.1103/PhysRevA.54.R1773}{\bibinfo {journal} {Phys. Rev. A}}\ }%
  \textbf{\bibinfo {volume} {54}},\ \bibinfo {pages} {R1773} (\bibinfo {year}
  {1996})%
  \bibAnnoteFile{NoStop}{Korsunsky1996}%
\bibitem{Suominen1996b}%
  \BibitemOpen
  \bibfield{author}{%
  \bibinfo {author} {\bibfnamefont{K.-A.}\ \bibnamefont{Suominen}},\ }%
  \bibfield{journal}{%
  \Doi{10.1088/0953-4075/29/24/008}{\bibinfo {journal} {J. Phys. B}}\ }%
  \textbf{\bibinfo {volume} {29}},\ \bibinfo {pages} {5981} (\bibinfo {year}
  {1996})%
  \bibAnnoteFile{NoStop}{Suominen1996b}%
\bibitem{Mastwijk1998}%
  \BibitemOpen
  \bibfield{author}{%
  \bibinfo {author} {\bibfnamefont{H.~C.}\ \bibnamefont{Mastwijk}}, \bibinfo
  {author} {\bibfnamefont{J.~W.}\ \bibnamefont{Thomsen}}, \bibinfo {author}
  {\bibfnamefont{P.}~\bibnamefont{van~der Straten}},\ and\ \bibinfo {author}
  {\bibfnamefont{A.}~\bibnamefont{Niehaus}},\ }%
  \bibfield{journal}{%
  \Doi{10.1103/PhysRevLett.80.5516}{\bibinfo {journal} {Phys. Rev. Lett.}}\ }%
  \textbf{\bibinfo {volume} {80}},\ \bibinfo {pages} {5516} (\bibinfo {year}
  {1998})%
  \bibAnnoteFile{NoStop}{Mastwijk1998}%
\bibitem{Glover2011}%
  \BibitemOpen
  \bibfield{author}{%
  \bibinfo {author} {\bibfnamefont{R.~D.}\ \bibnamefont{Glover}}, \bibinfo
  {author} {\bibfnamefont{J.~E.}\ \bibnamefont{Calvert}}, \bibinfo {author}
  {\bibfnamefont{D.~E.}\ \bibnamefont{Laban}},\ and\ \bibinfo {author}
  {\bibfnamefont{R.~T.}\ \bibnamefont{Sang}},\ }%
  \bibfield{journal}{%
  \Doi{10.1088/0953-4075/44/24/245202}{\bibinfo {journal} {J. Phys. B}}\ }%
  \textbf{\bibinfo {volume} {44}},\ \bibinfo {pages} {245202} (\bibinfo {year}
  {2011})%
  \bibAnnoteFile{NoStop}{Glover2011}%
\bibitem{Suominen1994}%
  \BibitemOpen
  \bibfield{author}{%
  \bibinfo {author} {\bibfnamefont{K.-A.}\ \bibnamefont{Suominen}}, \bibinfo
  {author} {\bibfnamefont{M.~J.}\ \bibnamefont{Holland}}, \bibinfo {author}
  {\bibfnamefont{K.}~\bibnamefont{Burnett}},\ and\ \bibinfo {author}
  {\bibfnamefont{P.~S.}\ \bibnamefont{Julienne}},\ }%
  \bibfield{journal}{%
  \Doi{10.1103/PhysRevA.49.3897}{\bibinfo {journal} {Phys. Rev. A}}\ }%
  \textbf{\bibinfo {volume} {49}},\ \bibinfo {pages} {3897} (\bibinfo {year}
  {1994})%
  \bibAnnoteFile{NoStop}{Suominen1994}%
\bibitem{Suominen1996a}%
  \BibitemOpen
  \bibfield{author}{%
  \bibinfo {author} {\bibfnamefont{K.~A.}\ \bibnamefont{Suominen}}, \bibinfo
  {author} {\bibfnamefont{K.}~\bibnamefont{Burnett}},\ and\ \bibinfo {author}
  {\bibfnamefont{P.~S.}\ \bibnamefont{Julienne}},\ }%
  \bibfield{journal}{%
  \Doi{10.1103/PhysRevA.53.R1220}{\bibinfo {journal} {Phys. Rev. A}}\ }%
  \textbf{\bibinfo {volume} {53}},\ \bibinfo {pages} {R1220} (\bibinfo {year}
  {1996})%
  \bibAnnoteFile{NoStop}{Suominen1996a}%
\bibitem{Suominen1998}%
  \BibitemOpen
  \bibfield{author}{%
  \bibinfo {author} {\bibfnamefont{K.-A.}\ \bibnamefont{Suominen}}, \bibinfo
  {author} {\bibfnamefont{Y.~B.}\ \bibnamefont{Band}}, \bibinfo {author}
  {\bibfnamefont{I.}~\bibnamefont{Tuvi}}, \bibinfo {author}
  {\bibfnamefont{K.}~\bibnamefont{Burnett}},\ and\ \bibinfo {author}
  {\bibfnamefont{P.~S.}\ \bibnamefont{Julienne}},\ }%
  \bibfield{journal}{%
  \Doi{10.1103/PhysRevA.57.3724}{\bibinfo {journal} {Phys. Rev. A}}\ }%
  \textbf{\bibinfo {volume} {57}},\ \bibinfo {pages} {3724} (\bibinfo {year}
  {1998})%
  \bibAnnoteFile{NoStop}{Suominen1998}%
\bibitem{Gallagher1989}%
  \BibitemOpen
  \bibfield{author}{%
  \bibinfo {author} {\bibfnamefont{A.}~\bibnamefont{Gallagher}}\ and\ \bibinfo
  {author} {\bibfnamefont{D.~E.}\ \bibnamefont{Pritchard}},\ }%
  \bibfield{journal}{%
  \Doi{10.1103/PhysRevLett.63.957}{\bibinfo {journal} {Phys. Rev. Lett.}}\ }%
  \textbf{\bibinfo {volume} {63}},\ \bibinfo {pages} {957} (\bibinfo {year}
  {1989})%
  \bibAnnoteFile{NoStop}{Gallagher1989}%
\bibitem{Julienne1991}%
  \BibitemOpen
  \bibfield{author}{%
  \bibinfo {author} {\bibfnamefont{P.~S.}\ \bibnamefont{Julienne}}\ and\
  \bibinfo {author} {\bibfnamefont{J.}~\bibnamefont{Vigu\'e}},\ }%
  \bibfield{journal}{%
  \Doi{10.1103/PhysRevA.44.4464}{\bibinfo {journal} {Phys. Rev. A}}\ }%
  \textbf{\bibinfo {volume} {44}},\ \bibinfo {pages} {4464} (\bibinfo {year}
  {1991})%
  \bibAnnoteFile{NoStop}{Julienne1991}%
\bibitem{Holland1994a}%
  \BibitemOpen
  \bibfield{author}{%
  \bibinfo {author} {\bibfnamefont{M.~J.}\ \bibnamefont{Holland}}, \bibinfo
  {author} {\bibfnamefont{K.-A.}\ \bibnamefont{Suominen}},\ and\ \bibinfo
  {author} {\bibfnamefont{K.}~\bibnamefont{Burnett}},\ }%
  \bibfield{journal}{%
  \Doi{10.1103/PhysRevLett.72.2367}{\bibinfo {journal} {Phys. Rev. Lett.}}\ }%
  \textbf{\bibinfo {volume} {72}},\ \bibinfo {pages} {2367} (\bibinfo {year}
  {1994})%
  \bibAnnoteFile{NoStop}{Holland1994a}%
\bibitem{Holland1994b}%
  \BibitemOpen
  \bibfield{author}{%
  \bibinfo {author} {\bibfnamefont{M.~J.}\ \bibnamefont{Holland}}, \bibinfo
  {author} {\bibfnamefont{K.-A.}\ \bibnamefont{Suominen}},\ and\ \bibinfo
  {author} {\bibfnamefont{K.}~\bibnamefont{Burnett}},\ }%
  \bibfield{journal}{%
  \Doi{10.1103/PhysRevA.50.1513}{\bibinfo {journal} {Phys. Rev. A}}\ }%
  \textbf{\bibinfo {volume} {50}},\ \bibinfo {pages} {1513} (\bibinfo {year}
  {1994})%
  \bibAnnoteFile{NoStop}{Holland1994b}%
\bibitem{Burnett1996}%
  \BibitemOpen
  \bibfield{author}{%
  \bibinfo {author} {\bibfnamefont{K.}~\bibnamefont{Burnett}}, \bibinfo
  {author} {\bibfnamefont{P.~S.}\ \bibnamefont{Julienne}},\ and\ \bibinfo
  {author} {\bibfnamefont{K.-A.}\ \bibnamefont{Suominen}},\ }%
  \bibfield{journal}{%
  \Doi{10.1103/PhysRevLett.77.1416}{\bibinfo {journal} {Phys. Rev. Lett.}}\ }%
  \textbf{\bibinfo {volume} {77}},\ \bibinfo {pages} {1416} (\bibinfo {year}
  {1996})%
  \bibAnnoteFile{NoStop}{Burnett1996}%
\bibitem{Piilo2006}%
  \BibitemOpen
  \bibfield{author}{%
  \bibinfo {author} {\bibfnamefont{J.}~\bibnamefont{Piilo}}, \bibinfo {author}
  {\bibfnamefont{E.}~\bibnamefont{Lundh}},\ and\ \bibinfo {author}
  {\bibfnamefont{K.-A.}\ \bibnamefont{Suominen}},\ }%
  \bibfield{journal}{%
  \Doi{10.1140/epjd/e2006-00147-6}{\bibinfo {journal} {Eur. Phys. J. D}}\ }%
  \textbf{\bibinfo {volume} {40}},\ \bibinfo {pages} {211} (\bibinfo {year}
  {2006})%
  \bibAnnoteFile{NoStop}{Piilo2006}%
\end{thebibliography}%

\end{document}